\documentclass[%
 reprint,
 amsmath,amssymb,
 aps,
]{revtex4-2}

\usepackage{graphicx}
\usepackage{dcolumn}
\usepackage{bm}
\usepackage{xcolor}

\begin{document}

\preprint{APS/123-QED}

\title{Superradiant droplet emission from a single hydrodynamic cavity near a reflective boundary}

\author{Konstantinos Papatryfonos}
\affiliation{Department of Mathematics, Massachusetts Institute of Technology} 

\author{Jemma W. Schroder}
\affiliation{Department of Mathematics, Massachusetts Institute of Technology.}

\author{Valeri Frumkin}
\affiliation{Department of Mechanical Engineering, Boston University.}

\begin{abstract}
Recent advances in manipulating droplet emissions from a thin vibrating fluid using submerged cavities, have introduced an innovative platform for generating hydrodynamic analogs of quantum and optical systems. This platform unlocks unique features not found in traditional pilot-wave hydrodynamics, inviting further exploration across varied physical settings to fully unravel its potential and limitations as a quantum analog. In this study, we explore how the recently reported phenomenon of hydrodynamic superradiance is affected when a single hydrodynamic cavity is taken to interact with a submerged reflective barrier. Our experimental findings reveal that the presence of a barrier near a cavity enhances its droplet emission rate, emulating the effect of a second cavity positioned at twice the distance.
Moreover, the system exhibits a sinusoidal modulation of the emission rate as a function of the distance between the cavity and its mirror image, echoing the characteristic superradiance signature observed in optical systems. These findings broaden our understanding of wave-particle duality in hydrodynamic quantum analogs and suggest new pathways for replicating quantum behaviors in macroscopic systems.       
\end{abstract}

\maketitle

\section{Introduction}

The field of hydrodynamic quantum analogs (HQA) presents a fascinating intersection where some classical fluid systems can behave analogously to corresponding quantum mechanical systems. Over the past two decades, a rapidly growing number of HQAs \citep{bush_hydrodynamic_2020} have been developed, aiming to build bridges between the tangible realm of fluid dynamics and the abstract quantum world. The most well-explored HQA system was discovered by Yves Couder, Emmanuel Fort, and colleagues in 2005 \citep{couder_single-particle_2006}, and it involves millimeter-sized droplets bouncing on a vibrating bath of the same fluid. Under specific experimental conditions, these bouncing droplets may self-propel along the surface of the bath, guided by a ‘pilot’ wave field of their own making, in many ways reminiscent of de Broglie's pilot-wave theory. The resulting object, called a walker, combines both wave and discrete particle properties, allowing it to replicate a plethora of quantum phenomena in fluid systems, including single-particle interference \citep{couder_single-particle_2006}, statistical projection \citep{saenz_statistical_2018}, superradiance \citep{frumkin_hydrodynamic_2023, papatryfonos2022}, emergent order \citep{saenz_emergent_2021}, interaction-free measurement \citep{frumkin_real_2022,frumkin_misinference_2023}, bipartite correlations in a static Bell setting \citep{Papatry_Bell}, and many more \citep{bush_state_2024, bush_perspectives_2024}. 

One limitation of the bouncing drops system is that it lacks mechanisms for particle creation and annihilation. Addressing this gap, recent work by Frumkin et al. \citep{frumkin_hydrodynamic_2023} has introduced a new hydrodynamic system endowed with such mechanisms, promising to generate a new class of HQAs. The system consisted of a shallow liquid layer interconnecting several deep circular cavities that were parametrically excited above their local Faraday threshold. The strength of the parametric forcing was then increased until the amplitude of the Faraday waves inside the cavities became comparable to their wavelength, at which point, the waves would spontaneously break leading to emission of droplets from the cavities via a process known as interfacial fracture. 
After emission, the droplets were reabsorbed back into the bulk liquid, and the process would repeat. 
Based on this hydrodynamic cavity emission system \citep{frumkin_hydrodynamic_2023}, the authors demonstrated a hydrodynamic version of superradiance, in which two cavities where placed in close proximity to one another, leading to droplets being emitted from the pair at a rate greater than twice that of a single cavity.
When the inter-cavity distance was varied between consecutive experiments, the authors observed a sinusoidal modulation of the droplet emission rate,
reminiscent of that observed in the behavior of cold-atom superradiance in quantum optics \citep{devoe_observation_1996, makarov_spontaneous_2003}. This sinusoidal modulation of the emission rate of the hydrodynamic cavities was shown to be a direct result of them being coupled via a common wavefield, created and shared by both cavities on the surface of the bath.

\begin{figure*} [t!]
\noindent \begin{centering}
\hspace*{-0.2cm}
\includegraphics[width=\textwidth]{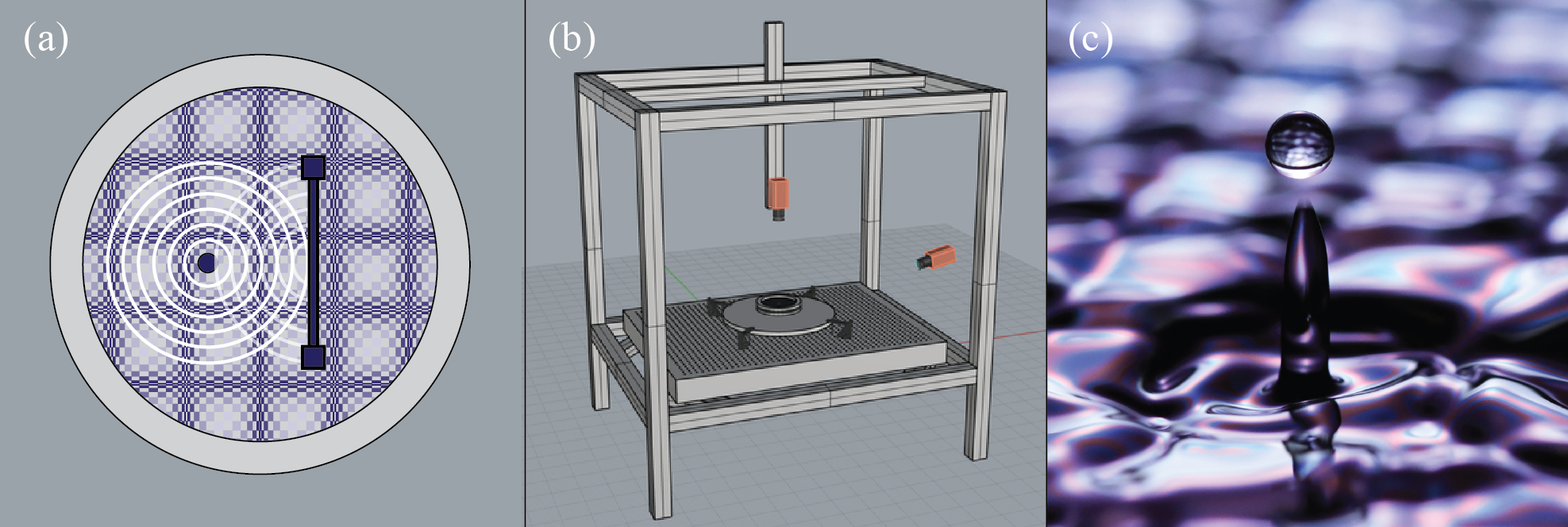}
\par\end{centering}
    \caption{The experimental setup. (a) A top view illustration of a circular bath filled with fluorinated oil, containing a circular cavity that emits concentric surface waves which propagate towards a submerged barrier that is 6 mm in width and 0.6 mm in depth. The distance between the barrier and the cavity was varied from $1.5$ to $4.5$ mm, in $0.5$ mm increments, and the steady-state emission rate was measured for each distance. The interaction between the reflected waves and the cavity affect the 
    (b) A perspective view of the experimental setup that includes a circular bath that is placed on an optical table and vertically oscillated by an electromagnetic shaker. (c) A snapshot of a droplet emission event.}
    \label{fig:1}
\end{figure*}

In the current study, we further advance this system by replacing one of the cavities with a submerged reflective boundary, drawing an analogy to the optical demonstration of interference effects produced by a single atom and its mirror image  \citep{eschner_light_2001}. 
Our experiments reveal a sinusoidal modulation of the emission rate as a function of the distance between the cavity and its 'mirror,' exhibiting features consistent with those of the two-cavity hydrodynamics case, and a generic signature of superradiance. 
Notably, while individual droplet emission events are unpredictable, the statistical behavior of the system remains robust, with well-defined emission probabilities that facilitate the demonstration of superradiance. 
The new experiments conducted here confirm the robustness of the statistical behavior of this new hydrodynamic quantum analog system, while demonstrating an effect that cannot be captured within the hydrodynamic pilot-wave framework. As such, this study underscores the added value of the hydrodynamic-cavity emission system, showcasing its potential as a complementary analog platform capable of introducing new features and expanding the scope of hydrodynamic quantum analogs.

\section{Results}

A complete analysis of the mechanism underlying the dynamics of this system, as well as a detailed description of the experimental setup used in these experiments, are provided in ref. \citep{frumkin_hydrodynamic_2023}. 
We briefly summarize here the main points. When a flat liquid surface is vibrated vertically at a fixed frequency, there exist two distinct stability thresholds associated with an increase in the vibration amplitude of the system. The first is known as the Faraday threshold $\gamma_F$, where the initially flat liquid surface destabilizes into a pattern of standing surface waves, known as Faraday waves. The second threshold, $\gamma_B$, occurs when the amplitude of the Faraday waves becomes comparable with their wavelength, at which point the waves spontaneously break, emitting droplets at random positions and times. These droplets may quickly re-integrate with the fluid, or remain bouncing on the surface depending on the fluid density, drop size, and vibration amplitude and frequency.
Importantly, in shallow liquids both $\gamma_F$ and $\gamma_B$ become highly sensitive to local depth, allowing the newly developed hydrodynamic-cavity emission system to access multiple regimes simultaneously by using topography with varying depth.  For example, in ref. \citep{frumkin_hydrodynamic_2023} the topography consisted of two circular cavities that served as deep regions operating above the fracture threshold while shallower regions remain below the fracture but above the Faraday threshold. In this regime, all emission events were confined to the regions above the deep cavities, while shallower regions allowed for inter-cavity coupling, resulting in the statistical behaviour that mimicked that of cold-atom superradiance in quantum optics \citep{devoe_observation_1996, makarov_spontaneous_2003}. 

In the current study, we replaced one of the cavities with a submerged barrier with the goal of reflecting the wavefield generated by a single cavity, similar to how an optical mirror reflects the electromagnetic field (see Fig. 1a). For the barrier we used a strip of plastic, 6 mm in width and 0.6 mm in depth, that was attached to the bottom plate (see Fig. 1(a)). The distance between the strip and the cavity was varied from $1.5$ to $4.5$ mm, in $0.5$ mm increments, and the steady-state emission rate was measured for each distance. To minimize any boundary effects, we ensured the strip’s end-to-end length was sufficiently long. The entire fluid bath was mounted on an optical table, and vertically vibrated using an electromagnetic shaker with forcing $F(t)=\gamma\cos(2\pi ft)$, where $\gamma$ denotes the peak vibrational acceleration and $f$ the frequency. A detailed description of the vibrational system (see Fig. 1b) that we used in order to maintain a constant vibrational acceleration amplitude to within $\pm 0.002$ g, is available in reference \citep{harris_generating_2015}. All experiments were conducted at a fixed driving frequency of $f=39$ Hz, close to the natural frequency of the cavities, as identified by an initial frequency sweep analysis that determined the maximal emission rate at a fixed amplitude of $1.75$ g. This specific amplitude facilitates a high droplet production rate while mitigating vibration-induced drainage of the cavities, which starts to appear at higher amplitudes. 

Droplet generation events were captured at 480 frames per second (fps) using a regular video camera (One Plus 6, 16MP camera, Sony IMX 519 sensor) placed beside the bath. An event was categorized as an emission event only when the droplet fully detached from the liquid thread. A still shot of a typical emission event is given in Fig 1c.  The emission events were manually counted, with each labeled according to its occurrence time. To visualize the wavefield, a semi-reflective mirror was angled at 45° between the bath and a charge-coupled device (CCD) camera mounted above the setup. Illumination was provided by a diffuse-light lamp oriented horizontally towards the mirror, generating bright regions that correspond to horizontal surfaces, extrema or saddle points. The hydrodynamic cavity was fashioned from a 6 mm deep acrylic plate with a centrally positioned 7 mm diameter hole, set atop a flat steel base. The optimal liquid depth in the shallow region around the cavity was established experimentally and its consistency between successive experiments, crucial for obtaining reliable statistics, is ensured by calibrating the liquid volume. Specifically, adding 20 ml of fluorinated oil produced a shallow bath depth of 0.76 $\pm$ 0.05 mm, verified by measurements taken before each experiment. 

We begin by characterizing the emission rate of a single hydrodynamic cavity in the absence of a barrier. Similar to Frumkin et al.,\citep{frumkin_hydrodynamic_2023} we find that individual droplet emission events are unpredictable, which we verify using a fast Fourier transform (FFT) analysis on the time series data. 
Specifically, by defining a binary drop generation function, where emission events are marked as 1 and non-events as 0, we show the absence of dominant Fourier components in the FFT analysis, underscoring the unpredictability of individual emission events. Despite this unpredictability, the steady state emission rate of a single cavity quickly converges to a fixed value, which we denote as $\Gamma_0$.
To insure that the Faraday pattern in the shallow layer outside the cavities does not influence the emission rate, we conducted repeated measurements of $\Gamma_0$, while repositioning the single cavity across various randomly-selected locations in the bath. These tests consistently yielded the same steady state rate of $\Gamma_0$, confirming the absence of any such effect in our experiments. This outcome aligns with our expectations, as the wavefield amplitude in the shallow region is significantly lower than that generated within the cavities. Additionally, these experiments verify that the influence of the bath's outer border on our observations is negligible.

Next, we introduce the barrier into the system, and measure the effect of its presence on the steady state emission rate.
The main results are summarized in Figure 2, which demonstrates the dependence of the anomalous emission rate, $\Gamma_N(d) = \left(\Gamma(d)-2\Gamma_0\right)/2\Gamma_0$ on the distance between the centre of the cavity and the reflective boundary, $d$. These results demonstrate that the emission rate $\Gamma_N(d)$ of the cavity at the presence of the reflective barrier is augmented with respect to the individual rate. Moreover, the rate increase is comparable to the one found in \citep{frumkin_hydrodynamic_2023} where superradiance by two cavities was demonstrated, when twice the distance between the cavity and the barrier is considered (or equivalently the distance between the cavity and its mirror image). The peak magnitude of the emission rate increase is found to be relatively smaller, reaching about 15 \% here for the cavity-barrier system, compared to 40 \% found for the two-cavities system.

Varying the distance between the cavity and the hydrodynamic barrier resulted in a sinusoidal modulation of the emission rate. Using linear regression, we fitted the function $\Gamma_N(d) = A\cos(k_f(d + h))$, where $A = 0.02$, $k_f = 0.91$, and $h = 0.62$. The fitted wavenumber corresponds to a wavelength $\lambda_f$ = $2\pi/k_f = 6.9$ mm, slightly longer than the experimentally measured wavelength of $\lambda=6.6\pm0.05$mm in the case of two cavities \citep{frumkin_hydrodynamic_2023}.  To estimate the vertical error bars, we employed a running average of the emission rate $\Gamma(d)$ defined as $S_d(T)=N/T$, where $N$ is the total number of emission events up to time $T$. As $T$ approaches infinity, $S_d(T)$ converges to the steady-state emission rate $\Gamma(d)$. The vertical error bars were then calculated as two standard deviations from the mean value of $S_d(T)$, representing the variability in the emission rate.  The horizontal error bars, reflect the mechanical precision error of the laser cutter used to fabricate the hydrodynamic cavity.

\begin{figure} [t!]
\noindent \begin{centering}
\hspace*{-0.5cm}
\includegraphics[width=21pc]{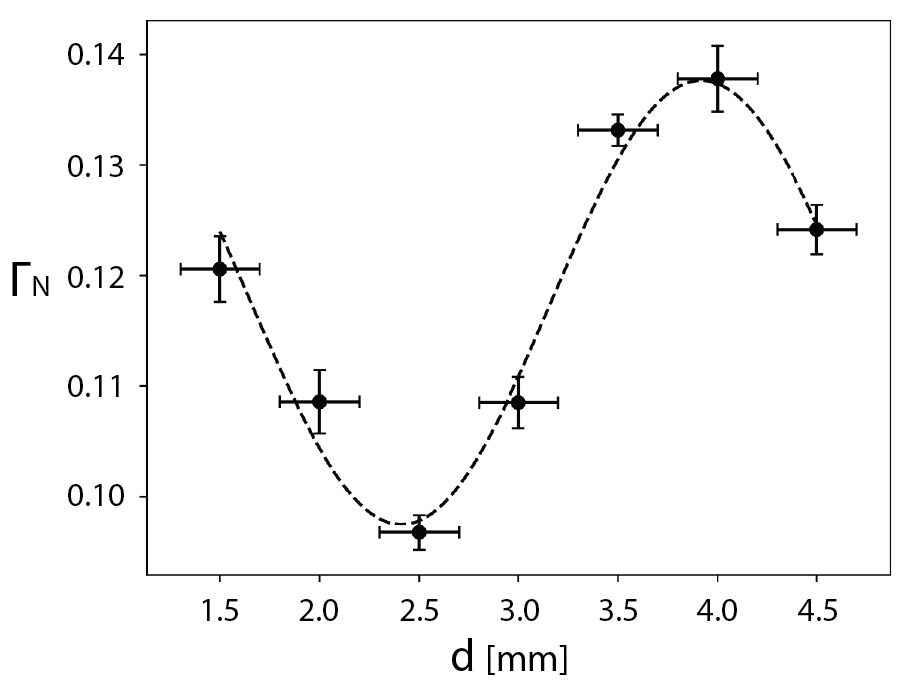}
\par\end{centering}
    \caption{Experimental data of the anomalous emission rate, $\Gamma_N(d) = \left(\Gamma(d)-2\Gamma_0\right)/2\Gamma_0$ (black dots) as a function of the distance between the centre of the cavity and the reflective boundary, $d$. Each data point represents an average over a time interval of $300$ seconds, corresponding to about $1000$ droplet emission events. The dashed curve represents $A\cos(k_f(d + h)) $, where $A = 0.0217$, $h = 0.6191$, and $k_f = 0.9100$  is the experimentally measured Faraday wave number in the vicinity of the cavities.}
    \label{fig:2}
\end{figure}

The sinusoidal modulation of the emission rates with distance is a characteristic signature of superradiance when the separation between the emitters, or the emitter and its mirror image, is comparable to the wavelength \citep{devoe_observation_1996, tanji-suzuki_chapter_2011, eschner_light_2001, frumkin_hydrodynamic_2023}.  The sinusoidal dependence that we observe here is analogous to the one shown by Eschner et al.\citep{eschner_light_2001}, in their demonstration of superradiance between a single atom and its mirror image in quantum optics.  
Our results, thus, further indicate that the emission rate of a single cavity is impacted by its own reflected wavefield, creating a similar effect as if a second cavity were placed at twice the distance.   

\section{Discussion}
In this work, we have demonstrated that the presence of a reflective boundary in the proximity of a parametrically excited hydrodynamic cavity may yield superradiant droplet emission, in a way analogous to atom-mirror systems in quantum optics.
In our system, superradiance arises from the fact that the excited cavity not only emits droplets, but also radiates a wavefield outwards \citep{frumkin_hydrodynamic_2023}. This wavefield couples the cavity to its surrounding objects, like other cavities or submerged barriers. The strength of this coupling strongly depends on their exact separation, particularly when this distance is on the order of the wavelength, a relationship that is also substantiated by the present study.
In quantum optics, light from an atom and its mirror image, that is, light from a single atom scattered into opposite directions, remains coherent and can therefore interfere \citep{eschner_light_2001}. This enables the use of mirrors to construct optical cavities and complex circuits, which allow control over many aspects of quantum technologies, from the emission rates and directionality in single-photon sources\citep{somaschi_near-optimal_2016}, to the engineering of topological nanophononic interface states using distributed Bragg reflectors\citep{rodriguez_topological_2023-1}. 
Analogously, in hydrodynamics, coherent surface waves reflecting off submerged barriers allow to control probabilities of droplet emission events. This opens many possibilities of using such analog systems to implement novel technologies, such as quantum-inspired computing, in a highly controllable and reliable environment.  

Finally, it is also worth noting that in this and previous studies \citep{frumkin_hydrodynamic_2023}, a high-density fluorinated oil has been used in order to facilitate rapid re-absorption of the emitted droplets back into the bulk liquid. This allowed avoiding extended drainage of the cavities, and prevented the emitted droplets from generating their own wavefields, which could interfere with the statistics needed to establish the emission rate of each cavity. However, if one uses other liquids such as silicon oil, the emitted droplets may not coalesce back into the bulk, and instead would continue bouncing on the surface of the vibrating liquid. In the correct parameter regime, these emitted droplets may form walkers that after emission would self-propel away from the cavity, thus serving as a direct link between the analog system presented here and the hydrodynamic pilot-wave system, providing a much richer source of analogs in future studies.


\vspace{0.5cm}
\subsection*{Acknowledgments} 
The authors gratefully acknowledge the financial support of the United States Department of State (V.F.) and
the European Union’s Horizon 2020 research and innovation programme under the Marie Sklodowska-Curie project EnHydro, grant agreement No 841417 (K.P.).

\subsubsection*{Competing interests}
The authors declare no competing interests.

\subsubsection*{Data and materials availability}
All data are available in the main text or the supplementary materials.

\end{document}